\documentclass[preprintnumbers,amsmath,amssymb]{revtex4}
\usepackage{epsfig}
\usepackage{color}
\usepackage{dsfont}
\usepackage[colorlinks,urlcolor=blue,linkcolor=blue,citecolor=red]{hyperref}
\begin{document}
\setlength{\voffset}{1.0cm}
\title{Duality study of chiral Heisenberg Gross-Neveu model in 1+1 dimensions}
\author{Michael Thies\footnote{michael.thies@gravity.fau.de}}
\affiliation{Institut f\"ur  Theoretische Physik, Universit\"at Erlangen-N\"urnberg, D-91058, Erlangen, Germany}
\date{\today}

\begin{abstract}
We consider a version of the Gross-Neveu model in 1+1 dimensions with discrete chiral and continuous flavor symmetry (isospin). 
In 2+1 dimensions, this model is known as chiral Heisenberg Gross-Neveu model. Spontaneous symmetry breaking and the
emergence of two massless and one massive scalar bosons are shown.
A duality to the Nambu--Jona-Lasinio model with isospin is exhibited, provided that the isovector pseudoscalar mean field is constrained
to a plane in isospin space. This enables us to find the phase diagram as a function of temperature, chemical potential and isospin
chemical potential as well as twisted kinks. A bare mass term acts quite
differently when added to this model as compared to other chiral variants of the Gross-Neveu model. 
\end{abstract}
\maketitle
\section{Introduction}
\label{sect1}
Four-fermion models in 1+1 dimensions can teach us a lot about strongly interacting relativistic systems. Well-known examples are the 
Gross-Neveu (GN) model \cite{1} with Z$_2\times$Z$_2$ chiral symmetry
($\psi \to \pm\gamma_5 \psi$),
\begin{equation}
{\cal L}_{\rm GN} = \bar{\psi} i \partial \!\!\!/ \psi + \frac{g^2}{2} \left(\bar{\psi}\psi \right)^2
\label{I1}
\end{equation}
and the Nambu--Jona-Lasinio (NJL) model \cite{2} with U(1)$\times$U(1) chiral symmetry ($\psi \to \exp\{i (\alpha + \beta \gamma_5)\}\psi$),
\begin{equation}
{\cal L}_{\rm NJL} = \bar{\psi} i \partial \!\!\!/ \psi + \frac{g^2}{2}\left[ (\bar{\psi}\psi)^2 + ( \bar{\psi} i \gamma_5 \psi )^2 \right].
\label{I2}
\end{equation}
If one includes isospin into the latter, one gets the NJL model with isospin (isoNJL) \cite{3} and non-Abelian SU(2)$\times$SU(2) chiral symmetry,
\begin{equation}
{\cal L}_{\rm isoNJL} = \bar{\psi} i \partial \!\!\!/ \psi + \frac{g^2}{2}\left[ (\bar{\psi}\psi)^2 + ( \bar{\psi} i \gamma_5 \vec{\tau} \psi )^2 \right].
\label{I3}
\end{equation}
In all three cases, one usually assumes that the fermions come in $N_c$ ``colors" ($\bar{\psi}\psi = \sum_{i=1}^{N_c} \bar{\psi}_i \psi_i$ etc.).
To leading order in the large $N_c$ limit \cite{4}, the models can then be solved explicitly using semiclassical methods.
Previous studies have uncovered a rich variety of fermion-antifermion and multifermion bound states, time dependent scattering problems, as well as 
non-trivial phase diagrams as a function of temperature and chemical potentials.

This brief survey suggests to add one more variant to this list which seems to have been forgotten so far.  
Starting from the GN model (\ref{I1}), let us introduce SU(2) isospin and replace $\bar{\psi}\psi$ in the four-fermion interaction by the corresponding isovector 
$\bar{\psi} \vec{\tau} \psi$,
\begin{equation}
{\cal L}_{\rm isoGN} = \bar{\psi} i \partial \!\!\!/ \psi + \frac{g^2}{2} \left(\bar{\psi} \vec{\tau} \psi \right)^2.
\label{I4}
\end{equation}
We thus arrive at the GN model with isospin (isoGN) featuring Z$_2 \times$Z$_2$ chiral symmetry and SU(2) flavor. 
As a matter of fact, in  2+1 dimensions this model is known in the condensed matter literature as ``chiral Heisenberg Gross-Neveu model" \cite{5}, presumably because
the interaction term is reminiscent of the spin-spin interaction in the Heisenberg model of magnetism. It has played a role in the context of the quantum Hall effect 
and graphene recently \cite{6,7,8}. An overview of the salient features of all four models is given in Table~\ref{tab1}. 
This shows in a compact way in which sense the isoGN model is complementary to the other three models listed.
The last two lines also give original references to the exact phase diagrams and soliton content of the models which cannot possibly be reviewed here due to lack
of space.

\begin{center}
\begin{table}
\begin{tabular}{|c|c|c|c|c|}
\hline
 & GN & isoGN & NJL & isoNJL  \\
\hline
color & U($N_c$) & U($N_c$) & U($N_c$) & U($N_c$) \\
flavor & 1 & SU(2) & 1 & SU(2) \\
chiral symmetry& Z$_2\times$Z$_2$ & Z$_2\times$Z$_2$ & U(1)$\times$U(1) & SU(2)$\times$SU(2) \\
vacuum manifold & $\pm 1$ & S$^2$ & U(1) & SU(2) \\
massless bosons & 0 & 2 scalars & 1 pseudoscalar & 3 pseudoscalars \\
massive bosons & 1 scalar & 1 scalar & 1 scalar & 1 scalar \\
phase diagram & \cite{9} & this work & \cite{10,11} & \cite{12} \\
solitons & \cite{13} & this work & \cite{14,15,16} & \cite{17,18} \\
\hline
\end{tabular}
\caption{Survey of four-fermion models with Lagrangians (\ref{I1}-\ref{I4})}
\label{tab1}
\end{table}
\end{center}

Let us mention that all of these models can also be amended by a bare fermion mass. When added to the Lagrangian, a Dirac mass term ($\delta {\cal L} = -m_0 \bar{\psi}\psi$) 
breaks chiral symmetry explicitly and renders the solution of the models more challenging.

The isoGN model is clearly less attractive from a phenomenological point of view. Nevertheless, we propose to analyze its large $N_c$ limit in 1+1 dimensions for theoretical and pedagogical reasons in this
work. Questions which immediately come to one's mind are: Does the model possess twisted kinks like the other GN variants, and how can one find them? What does the phase diagram look like, notably
regarding inhomogeneous phases? Can one find explicit time dependent scattering solutions, and what can be said about the integrability of the model?
We shall see that it takes little more than a duality to infer many physical properties of the isoGN model from previous results for the isoNJL model.
Thus this investigation also serves to illustrate the power of dualities in a novel context, see Refs.~\cite{19,20,21} for earlier applications of dualities to GN type models.
Finally, including a bare mass term has consequences different from all the other models and is also worth studying.

The plan of this paper is as follows. In Sect.~\ref{sect2}, we introduce our main tool, a duality between the isoGN model and a modified isoNJL model.
Sect.~\ref{sect3} deals with the Hartree-Fock (HF) vacuum and gap equation. Sect.~\ref{sect4} sketches the random phase approximation (RPA) and the meson spectrum.
Sect.~\ref{sect5} presents the full phase diagram as a function of temperature and chemical potentials. We also point out that any solution of the 
standard GN model generates a solution of the isoGN model with rigid isospin axis. In Sect.~\ref{sect6}, twisted kinks, i.e., solitonic multi-fermion bound states
interpolating between two different vacua, are constructed using duality. The bound state of two such kinks is exhibited and the composition law
for twist is interpreted geometrically. In Sect.~\ref{sect7}, we take a first glance at the massive isoGN model, whereas Sect.~\ref{sect8} finishes with a short
summary.
\section{Duality}
\label{sect2}
Consider the GN model with isospin, Lagrangian (\ref{I4}). The following symmetries can immediately be read off:  Discrete chiral symmetry $\psi \to \gamma_5 \psi$,
U(1) fermion number, U($N_c$) color, SU(2) isospin (in this context, both color and isospin are flavors). The discrete chiral symmetry is shared by the GN model (\ref{I1}).
The divergence of the axial currents confirms that isoscalar and isovector axial charges are not conserved, unlike the corresponding vector charges,
\begin{eqnarray}
\partial_{\mu} j^{\mu} & = &  \partial_{\mu}  \bar{\psi} \gamma^{\mu} \psi = 0
\nonumber \\
\partial_{\mu} j_5^{\mu} & = &  \partial_{\mu} \bar{\psi} \gamma^{\mu} \gamma_5 \psi =    -2 g^2 (\bar{\psi} i \gamma_5 \vec{\tau} \psi )\cdot (\bar{\psi} \vec{\tau} \psi )
\nonumber \\
\partial_{\mu} \vec{j}^{\,\mu} & = &   \partial_{\mu}\bar{\psi} \gamma^{\mu} \vec{\tau }\psi =   0
\nonumber \\
\partial_{\mu} \vec{j}_5^{\,\mu} & = &  \partial_{\mu}    \bar{\psi} \gamma^{\mu} \gamma_5 \vec{\tau} \psi =  -2 g^2 (\bar{\psi} i \gamma_5 \psi ) (\bar{\psi} \vec{\tau} \psi ).
\label{A1}
\end{eqnarray}
Next we turn to the subject of duality. The authors of Refs.~\cite{20,21} have noted a kind of duality inside the isoNJL model. It amounts to the unitary  
transformation \cite{12}
\begin{equation}
{\rm U}_{\rm dual} = i \tau_3 P_L + i \tau_1 P_R, \quad P_{R,L} = \frac{1 \pm \gamma_5}{2}.
\label{A2}
\end{equation}
It acts as follows on the bilinears relevant for the isoNJL model with baryonic, isospin and axial isospin chemical potentials
\begin{eqnarray}
\bar{\psi} i \gamma_5 \tau_1 \psi & \leftrightarrow & \bar{\psi} i \gamma_5 \tau_3 \psi
\nonumber \\
\bar{\psi} i \gamma_5 \tau_2 \psi & \leftrightarrow & \bar{\psi} \psi
\nonumber \\
\psi^{\dagger} \tau_3 \psi & \leftrightarrow& - \psi^{\dagger} \gamma_5 \tau_3 \psi.
\label{A3}
\end{eqnarray}
This enables one to map mean field solutions involving only ``neutral" condensates ($S,P_3$) onto solutions involving only
``charged" condensates ($P_1 \pm i P_2$). Here,
\begin{equation}
S = - g^2 \langle \bar{\psi} \psi \rangle, \quad \vec{P} = - g^2 \langle \bar{\psi} i \gamma_5 \vec{\tau} \psi \rangle.
\label{A4}
\end{equation}
We propose a different transformation relating the isoGN model to a truncated version of the isoNJL model.
At first glance, the isoNJL model and the isoGN model cannot be dual to each other since they have different numbers of interaction terms or condensates.
Suppose that we only allow pseudoscalar isovector condensates $\vec{P}$ living in a certain plane in isospin space. Then a duality becomes viable, 
since both models have 3 condensates. This is also potentially interesting, relating a subset of known HF solutions of the isoNJL model to novel 
solutions of the isoGN model.
To this end, define the canonical transformation
\begin{equation}
\psi_L \to \tau_1 \psi_L, \quad \psi_R \to \psi_R
\label{A5}
\end{equation}
implemented by the unitary operator
\begin{equation}
{\rm T} = \tau_1 P_L + P_R = {\rm T}^{\dagger}, \quad {\rm T}^2=1.
\label{A6}
\end{equation}
By a global isospin rotation, $\tau_1$ could be rotated into any other component of $\vec{\tau}$, but we shall stick to the choice (\ref{A6}) 
for notational simplicity. 
${\rm T}$ acts as follows on the relevant Dirac- and isospin matrices
\begin{eqnarray}
{\rm T} \gamma^0  &  = &  \gamma^0 \tau_1 {\rm T}
\nonumber \\
{\rm T} i \gamma^1 \tau_1  & = & i \gamma^1 {\rm T}
\nonumber \\
{\rm T} i \gamma^1 \tau_2  & = & - \gamma^0 \tau_3 {\rm T}
\nonumber \\
{\rm T} i \gamma^1 \tau_3  & = & \gamma^0 \tau_2 {\rm T}
\nonumber \\
{\rm T} \tau_3  & = & \gamma_5 \tau_3 {\rm T}.
\label{A7}
\end{eqnarray}
This implies a number of dualities between bilinears (remember that $\gamma_5=\gamma^0\gamma^1$ in 1+1 dimensions)
\begin{eqnarray}
\bar{\psi} \psi & \leftrightarrow &    \bar{\psi} \tau_1 \psi 
\nonumber \\
\bar{\psi} i \gamma_5 \tau_1 \psi & \leftrightarrow &  \bar{\psi} i \gamma_5 \psi
\nonumber \\
\bar{\psi} i \gamma_5 \tau_2  \psi & \leftrightarrow & - \bar{\psi} \tau_3  \psi
\nonumber \\
\bar{\psi} i \gamma_5 \tau_3 \psi & \leftrightarrow &  \bar{\psi}  \tau_2 \psi
\nonumber \\
\psi^{\dagger} \tau_3 \psi & \leftrightarrow & \psi^{\dagger} \gamma_5 \tau_3 \psi .
\label{A8}
\end{eqnarray}
The double arrow reflects the fact that ${\rm T}^2=1$. These relations show that the isoNJL model (\ref{I3}) is dual to the Lagrangian
\begin{equation}
\tilde{\cal L}_{\rm isoNJL} = \bar{\psi} i \partial \!\!\!/ \psi + \frac{g^2}{2} \left[   \left(\bar{\psi} \vec{\tau} \psi \right)^2 + \left(\bar{\psi} i \gamma_5 \psi \right)^2\right].
\label{A9}
\end{equation}
In other words, we are allowed to swap scalar and pseudoscalar couplings in the two interaction terms, or, equivalently, isoscalar and isovector couplings.
Although distinct at first glance, ${\cal L}_{\rm isoNJL}$ of (\ref{I3}) and $\tilde{\cal L}_{\rm isoNJL}$ of (\ref{A9}) are just two ways of describing the same physics.
In order to arrive at ${\cal L}_{\rm isoGN}$, Eq.~(\ref{I4}), we
have to get rid of the pseudoscalar term in (\ref{A9}). Eq.~(\ref{A8}) tells us that we then have to start from the isoNJL Lagrangian, but omitting the term
$\sim (\bar{\psi}i \gamma_5 \tau_1\psi)^2$ (the 1-component is singled out by our choice of $\tau_1$ in the definition of ${\rm T}$).
The upshot is that the isoGN model (\ref{I4}) is dual to the following truncated version of the isoNJL model
\begin{equation}
\tilde{\cal L}_{\rm isoGN} = \bar{\psi} i \partial \!\!\!/ \psi + \frac{g^2}{2}\left[ (\bar{\psi}\psi)^2 + ( \bar{\psi} i \gamma_5 \tau_2 \psi )^2 + ( \bar{\psi} i \gamma_5 \tau_3 \psi )^2\right].
\label{A10}
\end{equation}
This duality will allow us to infer the yet unknown phase diagram of the isoGN model at finite chemical potential and 
isospin chemical potential from the known phase diagram of the isoNJL model at finite chemical potential and axial isospin chemical potential, without any additional effort.
It will also be useful for constructing solitonic multi-fermion bound states. 

A last remark on duality is in order. In the case of the isoNJL model, the unitary transformation U$_{\rm dual}$ (\ref{A2}) was an element of the chiral symmetry group SU(2)$\times$SU(2).
Without chemical potentials, it cannot have any physical effect since it does not matter which vacuum one picks in the case of spontaneous symmetry breaking (SSB). With chemical potentials
the situation is different because isospin and axial isospin chemical potentials are interchanged \cite{21}. In the present case, ${\rm T}$ does not belong to the symmetry group of the isoGN model,
but relates two seemingly distinct field theories. In this respect, the situation is more like the original example of a duality where four-fermion models with either Cooper pairing or chiral symmetry breaking have been related
\cite{19}. In that case, the duality was only recognized after both field theoretic models had already been solved independently \cite{22}. Here we shall take advantage of the fact that we have identified the
duality before solving one of the two models involved, namely the isoGN model.

\section{Vacuum}
\label{sect3}

Mean field theory for fermions means HF or time dependent Hartree-Fock (TDHF) in a relativistic setting. In the case of the isoGN model, the Lagrangian (\ref{I4})
gives rise to the TDHF equation
\begin{equation}
\left( i \partial \!\!\!/ - \vec{S}\cdot \vec{\tau} \right) \psi  =  0,
\quad
\vec{S} =  - g^2 \langle \bar{\psi} \vec{\tau} \psi \rangle 
\label{B1}
\end{equation} 
where the 2nd part is the self-consistency condition.
The corresponding stationary Schr\"odinger equation is
\begin{equation}
H\psi =\left( - i \gamma_5  \partial_x + \gamma^0 \vec{S}\cdot \vec{\tau} \right) \psi = E \psi.
\label{B2}
\end{equation}
To find the vacuum, we look for homogeneous solutions $\vec{S}=$const. which break SU(2) isospin and the discrete chiral symmetry spontaneously. 
It is trivial to diagonalize $H$ with constant $\vec{S}$ and the substitution $\partial_x \to ik$. The eigenvalues are $\pm \sqrt{k^2+M^2}$ (two times degenerate each)
with $M = |\vec{S}|$. The vacuum manifold is spanned by the real 3-vector $\vec{S}$ with fixed length, i.e., it is a 2-sphere $S^2$.
Global isospin rotations map one vacuum onto another one. If the order parameter minimizing the vacuum energy density does not vanish, we have SSB of the global SU(2) symmetry.
At the mean field level considered here, we then expect a pair of scalar (would-be) Goldstone bosons and a massive scalar meson, even in 1+1 dimensions.

We choose the isospin frame such that $\vec{S}$ points into the 1--direction.
If we then invoke the duality transformation ${\rm T}$, we come back to the free theory
with two flavors of massive Dirac fermions ($M=|S_1|$). The only remnant of the interactions is the 
self-consistency condition 
\begin{equation}
\vec{S}  =  -  g^2 \langle \bar{\psi} \vec{\tau} \psi \rangle .
\label{B3}
\end{equation}
Using a momentum cutoff $\Lambda/2$, it yields the gap equation in the form
\begin{equation}
1- \frac{2N_cg^2}{\pi} \ln \frac{\Lambda}{M} = 0.
\label{B4}
\end{equation}
The factor of 2 as compared to the original GN model is due to isospin (doubling of the total number of flavors, $N=2N_c$) and was already encountered in the isoNJL model \cite{12}.
The vacuum energy density per flavor coincides with that of the standard GN model,
\begin{equation}
\frac{{\cal E}_{\rm vac}}{2 N_c} = - \frac{M^2}{4\pi}.
\label{B5}
\end{equation}
The divergent energy density of the symmetric vacuum ($M=0$) has been subtracted as usual, so that the negative value indicates that symmetry breaking
is favored energetically. The value of $M$ is arbitrary, since the Lagrangian does not possess any scale, and can be set equal to 1.
All the well-known phenomena related to renormalization (asymptotic freedom, dimensional transmutation) are the same as in the standard GN model.

\section{Meson spectrum}
\label{sect4}

The meson spectrum of fermion-antifermion bound states can be inferred from small fluctuations around the HF vacuum. The appropriate machinery is 
the relativistic form of the RPA. Since it is fairly standard and we follow closely similar calculations in previous
works on GN type models \cite{10,23,24}, here we give only the principal definitions and sketch the main steps.
The central quantity is the one-body density matrix, expanded around the vacuum expectation value,
\begin{equation}
Q(x,y)  =   \rho(x-y)+ \frac{1}{\sqrt{N_c}} \tilde{Q}(x,y).
\label{C1}
\end{equation}
This $4 \times 4$ matrix (Dirac and isospin indices) is decomposed in terms of vacuum eigenspinors, where only the following pieces survive in the large $N_c$ limit,
\begin{equation}
\tilde{Q}(k',k)  =  u_a(k')v_b^{\dagger}(k) Q_{ab}^{12}(k',k)  + v_a(k')u_b^{\dagger}(k) Q_{ab}^{21}(k',k) .
\label{C2}
\end{equation}
Here, $u$ and $v$ denote positive and negative energy spinors, respectively, and the labels ${a,b}$ refer to isospin.
To leading order in $1/N_c$, the equation of motion for $Q$ is solved automatically by choosing vacuum spinors. Linearizing the equations in $\tilde{Q}$ is nothing
but the RPA. The meson spectrum can be obtained by sandwiching the bilinear fermion operator $\tilde{Q}$ between vacuum and one-meson states of momentum $P$,
\begin{eqnarray}
\langle P | \tilde{Q}_{ab}^{21}(k',k) |{\rm vac}\rangle &=& 2\pi \delta(P-k+k')X_{ab}(P,k),
\nonumber \\
\langle P | \tilde{Q}_{ab}^{12}(k',k) |{\rm vac}\rangle &=& 2\pi \delta(P-k+k')Y_{ab}(P,k).
\label{C3}
\end{eqnarray}
After the dust has settled, the RPA equations assume the form 
\begin{eqnarray}
X_{ab}(P,k) & = & - Ng^2 \frac{\bar{v}_{k-P,a}\tau^c u_{k,b}}{{\cal E}(P) - E(k-P,k)} Z^c(P),
\nonumber \\
Y_{ab}(P,k) & = & Ng^2  \frac{\bar{u}_{k-P,a}\tau^c v_{k,b}}{{\cal E}(P) + E(k-P,k)} Z^c(P),
\nonumber \\
Z^c(P) & = & \int \frac{dk'}{2\pi} \left[ \bar{v}_{k'b} \tau^c u_{k'-P,a} Y_{ab}(P,k')  + \bar{u}_{k'b} \tau^c v_{k'-P,a} X_{ab}(P,k')\right],
\label{C4}
\end{eqnarray}
reflecting the separable form of the kernel.
We have used the notation
\begin{eqnarray}
E(k',k) & = & E_{k'}+E_k, \quad  E_k=\sqrt{M^2+k^2},
\nonumber \\
{\cal E}(P) & =  & \sqrt{{\cal M}^2+ P^2}
\label{C5}
\end{eqnarray}
where ${\cal M}$ is the meson mass.
Eq.~(\ref{C4}) is a homogeneous linear system 
\begin{equation}
Z^c(P)  =  M^{cd}(P)Z^d(P)
\label{C6}
\end{equation}
with the matrix
\begin{eqnarray}
M^{cd}(P) & = & Ng^2 \int \frac{dk} {2\pi}  \left[ \frac{(\bar{v}_{k,b}\tau^c u_{k-P,a)})(\bar{u}_{k-P,a}\tau^d v_{k,b})}{{\cal E}(P)+E(k-P,k)} 
  - \frac{(\bar{u}_{k,b}\tau^c v_{k-P,a})(\bar{v}_{k-P,a}\tau^d u_{k,b})}{{\cal E}(P)-E(k-P,k)} \right]
\nonumber \\
& = & Ng^2 \int \frac{dk}{2\pi} m^{cd}(k,P).
\label{C7}
\end{eqnarray}
Upon working out the integrand $m^{cd}$ in Eq.~(\ref{C6}), we find that it is diagonal in isospin with ($k'=k-P$)
\begin{eqnarray}
m^{11}(k,P) & = & m^{22}(k,P)= \left( \frac{1}{E_k} + \frac{1}{E_{k'}} \right)  \frac{P^2-E^2(k',k)}{{\cal E}^2(P)-E^2(k',k)},
\nonumber \\
m^{33}(k,P) & = & \left( \frac{1}{E_k} + \frac{1}{E_{k'}} \right)  \frac{4M^2+P^2-E^2(k',k)}{{\cal E}^2(P)-E^2(k',k)}.
\label{C8}
\end{eqnarray} 
The first two entries reduce Eq.~(\ref{C6}) to the vacuum gap equation for ${\cal M}=0$, the third one for ${\cal M}=2M$. The covariant energy-momentum 
relation for the mesons is manifest. As expected, there are two massless ``would-be" Goldstone bosons 
matching the number of flat directions of the $S^2$ vacuum manifold and one massive scalar meson, the radial excitation. 
The massive meson has mass $2M$. Incidentally, the same marginally bound state has been found in the other variants of the GN model, Eqs.~(\ref{I1},\ref{I2},\ref{I3}).

\section{Phase diagram and HF solutions with fixed isospin direction}
\label{sect5}

\begin{figure}
\begin{center}
\epsfig{file=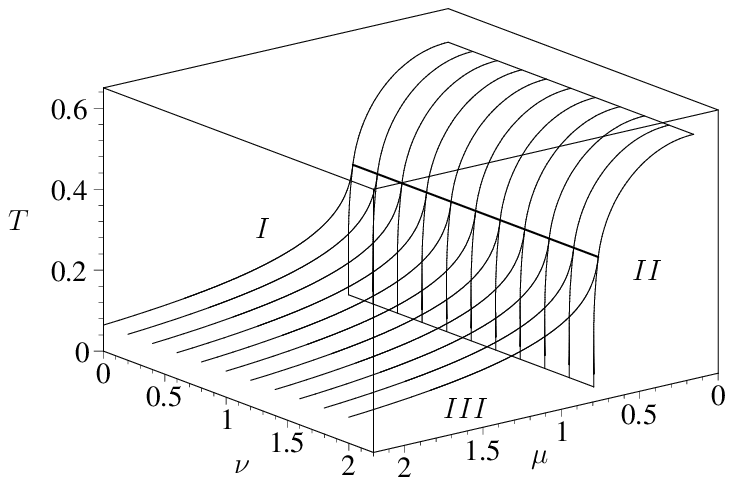,width=8cm,angle=0}
\caption{Full phase diagram of the massless isoGN model (units $M=1$). $I$) Chirally restored phase, $II$) homogeneous, massive phase, $III$) soliton crystal. 
The order parameter does not depend on $\nu$ and coincides with that of the GN model. Adapted from Ref.~\cite{12}.}
\label{fig1}
\end{center}
\end{figure}

We first show how to deduce the phase diagram of the isoGN model from that of the isoNJL model using duality. If we turn to thermal HF theory, two new aspects come
into the picture: First, the HF equation in canonical form is amended by fermionic ($\mu$) and isospin ($\nu$) chemical potentials,
\begin{equation}
\left( -i \gamma_5 \partial_x - \mu - \nu \tau_3 + \gamma^0 \vec{S} \cdot \vec{\tau} \right) \psi = E \psi.
\label{D1}
\end{equation}
Secondly, the self-consistency condition now involves thermal rather than ground state expectation values,
\begin{equation}
\vec{S} = - g^2 \langle \bar{\psi} \vec{\tau} \psi \rangle_{\rm therm}= - g^2 \sum_{\alpha} \bar{\psi}_{\alpha} \vec{\tau} \psi_{\alpha} \frac{1}{e^{\beta E_{\alpha}}+1}, \quad \beta=1/T.
\label{D2}
\end{equation}
Applying the duality transformation $\psi={\rm T}\phi$ to (\ref{D1}) yields
\begin{equation}
\left[ -i \gamma_5 \partial_x - \mu - \nu \gamma_5 \tau_3 + \gamma^0 S_1 +  i \gamma^1 \left( S_2 \tau_3-S_3 \tau_2 \right) \right] \phi = E \phi
\label{D3}
\end{equation}
with
\begin{equation}
S_1 = - g^2 \langle \bar{\phi} \phi \rangle_{\rm therm}, \quad S_2 =  - g^2 \langle \bar{\phi} i \gamma_5 \tau_3 \phi \rangle_{\rm therm}, \quad
S_3 = g^2 \langle \bar{\phi} i \gamma_5 \tau_2 \phi \rangle_{\rm therm}.
\label{D4}
\end{equation}
The problem has thus been mapped onto the HF equation for the isoNJL model with mean fields
\begin{equation}
S=S_1, \quad P_1=0, \quad P_2=-S_3, \quad P_3=S_2, 
\label{D5}
\end{equation}
fermion chemical potential $\mu$, vanishing isospin chemical potential and axial isospin chemical potential $\nu$.
The solution to this problem can be taken over from Ref.~\cite{12} simply by switching off the isospin chemical potential.
The resulting picture of the phase boundaries in ($\mu,\nu,T$)-space of the isoGN model is indistinguishable from that of the isoNJL model and reproduced in Fig.~\ref{fig1}.
The order parameters are different though. In the isoNJL model, the order parameter could be factorized as
\begin{equation}
S(\mu,\nu,\nu_5,T) = S_{\rm GN}(\mu,T)e^{2i\nu x}
\label{D6}
\end{equation}
and had no dependence on $\nu_5$. Since the isospin chemical potential $\nu$ of the isoGN model corresponds to $\nu_5$ of the isoNJL model,
the order parameter of the isoGN model reduces to that of the GN model \cite{9},
\begin{equation}
S_1(\mu,\nu,T)  = S_{\rm GN}(\mu,T) , \quad S_2 = S_3 = 0.
\label{D7}
\end{equation}
The only place where the isospin chemical potential shows up is in the value of the thermodynamic potential, namely
\begin{equation}
\left. \frac{{\cal V}_{\rm eff} (\mu,\nu,T)}{2N_c} \right|_{\rm isoGN} = \left. \frac{{\cal V}_{\rm eff}(\mu,T)}{N_c} \right|_{\rm GN} - \frac{\nu^2}{2\pi}.
\label{D8}
\end{equation}

The thermodynamic ground state is one example of a HF solution where the order parameter has a fixed direction in isospin space while depending on $x$. 
As a matter of fact, any HF or TDHF solution of the GN model generates a corresponding solution of the isoGN model with frozen isospin direction.
This can be seen as follows. For simplicity, let us look for mean field solutions of the isoGN model with $S_3 \neq 0$ only. In that case the TDHF problem reduces
to that of the standard GN model (\ref{I1}) for isospin up and a $\gamma_5$-transformed copy thereof for isospin down ($S$ changes sign),
\begin{equation}
\vec{S} \cdot \vec{\tau} = S_{\rm GN} \tau_3.
\label{D9}
\end{equation}
Here, $S_{\rm GN}$ is a self-consistent mean field of the standard GN model with $N=2 N_c$ flavors.
Indeed, the Schr\"odinger equation 
\begin{equation}
H\psi  = \left( \begin{array}{cc} i \partial_x & S_{\rm GN} \tau_3 \\ S_{\rm GN} \tau_3 & -i \partial_x \end{array} \right)\psi = i \partial_t \psi
\label{D10}
\end{equation}
admits the following solutions for isospin up/down states in terms of solutions of the GN model,
\begin{equation}
\psi_I = \left( \begin{array}{c} \psi_L \\ 0 \\ \psi_R \\ 0 \end{array} \right)_{\rm GN}, \quad \psi_{II} = \left( \begin{array}{c} 0 \\  - \psi_L \\ 0 \\ \psi_R  \end{array} \right)_{\rm GN}.
\label{D11}
\end{equation}
The matrix elements for single particle levels entering the self-consistency conditions become
\begin{eqnarray}
\bar{\psi}_I \tau_{1,2} \psi_I & = & \bar{\psi}_{II} \tau_{1,2} \psi_{II} = 0 
\nonumber \\
\bar{\psi}_I \tau_3 \psi_I & = & \bar{\psi}_{II} \tau_3 \psi_{II}  =  (\bar{\psi}\psi )_{\rm GN}.
\label{D12}
\end{eqnarray}
The first line is trivial since the expectation value of $\tau_{1,2}$ in an eigenstate of $\tau_3$ vanishes. These identities are sufficient to 
prove self-consistency by summing over all occupied states. Thus, all soliton solutions of the GN model can be adapted to the isoGN model, including the crystal solution and multi-soliton
bound and scattering states. As in the case of the vacuum, the two isospin states give identical contributions to the (isovector) condensate. The resulting
factor of 2 from the two copies accounts for $N=2 N_c$ in the gap equation. 

There is no reason to expect that this class of special solutions with fixed isospin direction exhausts all possibilities. This raises the question about solutions
with varying isospin direction to be addressed in the following section.

\section{Twisted kinks}
\label{sect6}

Here the duality becomes particularly useful. The dual Lagrangian (\ref{A10}) corresponds to the SU(2)$\times$SU(2) symmetric isoNJL model
minus the interaction term $\sim (\bar{\psi} i \gamma_5 \tau_1 \psi)^2$ for our choice of the isospin frame. Any HF or TDHF solution of the isoNJL model with identically 
vanishing $P_1$ can thus be used to generate a solution of the isoGN model.

Consider the twisted kink at rest of the isoNJL model \cite{18}. Vacua in the isoNJL model correspond to constant SU(2) matrices. The twisted kink interpolates between the vacua $\Delta_-=M$ at $x\to - \infty$ 
and $\Delta_+ = M \exp \left( -2i \theta \vec{n}\cdot \vec{\tau} \right)$ at  $x \to \infty$. Here, $\theta$ is called the twist angle and $P_1=0$ holds on condition that the unit vector $\vec{n}$ lies
in the (2,3)-plane,
\begin{equation}
\vec{n} = \left( \begin{array}{c} 0 \\ - \sin \beta \\ \cos \beta \end{array} \right).
\label{E1}
\end{equation}
In the representation 
\begin{equation}
\gamma^0 = \sigma_1, \quad \gamma^1 = i \sigma_2, \quad \gamma_5 = \gamma^0 \gamma^1 = - \sigma_3,
\label{E2}
\end{equation}
the Hamiltonian of the isoNJL model assumes the 4$\times$4 matrix form 
\begin{equation}
H = \left( \begin{array}{cc} i \partial_x & \Delta^{\dagger} \\ \Delta & -i \partial_x \end{array} \right) 
\label{E3}
\end{equation}
with the twisted kink potential $\Delta$ given by 
\begin{equation}
\Delta  =   \frac{\Delta_- + V \Delta_+}{1+V}  =  S-i \vec{P} \cdot \vec{\tau}, \quad
V = e^{2Mx\sin \theta}
\label{E4}
\end{equation}
Inserting $\Delta_{\pm}$, we read off
\begin{eqnarray}
S & = & M \left( \frac{1 + \cos (2\theta) V}{1+V} \right),
\nonumber \\
\vec{P} & = & M \left( \frac{\sin (2\theta) V}{1+V} \right) \vec{n}.
\label{E5}
\end{eqnarray}
After these preparations taken from \cite{18}, we now invoke duality. The dual twisted kink of the isoGN model will be characterized by the
hermitean potential $\Delta=\vec{S} \cdot \vec{\tau}$ with 
\begin{equation}
\vec{S} = \left( \begin{array}{c} S_1 \\ S_2 \\ S_3 \end{array} \right) = \left( \begin{array}{c} S \\ P_3 \\ - P_2 \end{array} \right)
\label{E6}
\end{equation}
where the last entries can be taken over literally from the isoNJL model, Eq.~(\ref{E5}). The twisted kink can again be cast into a form similar to (\ref{E4}), but now the
asymptotic vacua $\Delta_{\pm}= \vec{S}_{\pm} \cdot \vec{\tau}$ are elements of the su(2) Lie algebra rather than the SU(2) group,
\begin{equation}
\vec{S}  =   \frac{\vec{S}_- + V \vec{S}_+}{1+V},
\quad
\vec{S}_-  =  M \left( \begin{array}{c} 1 \\ 0 \\ 0 \end{array} \right) , \quad \vec{S}_+  = M \left( \begin{array}{c} \cos (2\theta) \\ \cos \beta \sin (2 \theta) \\ \sin \beta \sin (2 \theta) \end{array} \right).
\label{E7}
\end{equation}
Up to global isospin rotations, this is the most general twisted kink of the isoGN model. In isospin space, it
interpolates between the point $M$ on the 1-axis and an arbitrary point on the vacuum manifold $S^2$ of radius $M$. By an isospin rotation, we can transform this object into a kink interpolating between 
two different points $\vec{S}_{\pm}$ on the sphere, provided that the angle between ($\vec{S}_-, \vec{S}_+$) is the same. This angle is twice the twist angle $\theta$, an intrinsic property of the kink, and enters the scalar interpolating function $V$ as seen
in Eq.~(\ref{E4}). We remind the reader that the original twisted kink was constructed by Shei in the NJL model \cite{14} where its potential connects two points on the vacuum manifold, a circle of radius $M$, along 
a straight line segment (``chord soliton"). What is the trajectory traced out in isospin space by the twisted kink of the isoGN model? The kink (\ref{E7}) can equivalently be represented as 
\begin{equation}
\vec{S} = \vec{S}_- + \frac{V}{1+V} \left( \vec{S}_+ - \vec{S}_- \right)
\label{E8}
\end{equation} 
showing that it also follows a straight line segment, now connecting two points on the 2-sphere.
The interpolating function is a smooth, kink-like function rising from 0 to 1 as $x$ goes from $-\infty$ to $+ \infty$, 
\begin{equation}
\frac{V}{1+V} = \frac{1+ \tanh (M x \sin \theta)}{2}
\label{E9}
\end{equation}
The name kink is justified by the shape of this function, whereas twist refers to the asymptotic vacua, i.e., the vectors $\vec{S}_{\pm}$. The two are related in that the 
twist angle (half the angle between $\vec{S}_-$ and $\vec{S}_+$) also determines the steepness of the kink function (\ref{E9}).
All other details about the twisted kink (spinors, fermion number, proof of self-consistency, evaluation of mass) can be skipped here since they have been fully discussed in the dual theory.  Fermion density has the same
meaning in both models, but isospin density changes by a factor of $\gamma_5$ due to the duality transformation ${\rm T}$, see Eq.~(\ref{A8}). 
Let us just mention that the mass of the twisted isoGN kink is the same as that in the dual model,
\begin{equation}
M_{\rm kink} = \frac{2 N_c M \sin \theta}{\pi}.
\label{E10}
\end{equation}

\begin{figure}
\begin{center}
\epsfig{file=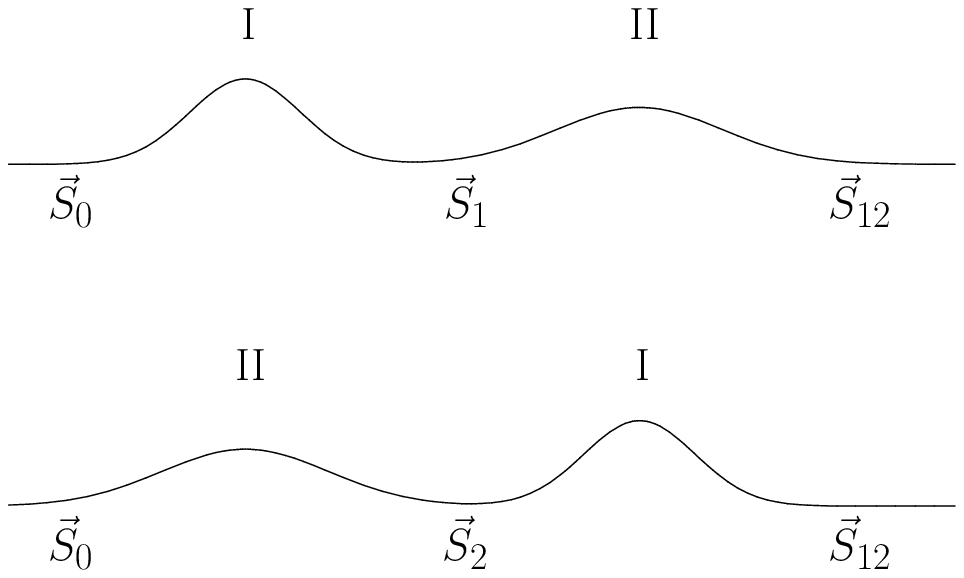,width=6cm,angle=0}
\caption{Schematic illustration of kink-kink bound states with two different spatial configurations, serving to explain the basis of the geometrical
composition of twists, Eqs.~(\ref{E16},\ref{E17}).}
\label{fig2}
\end{center}
\end{figure}

As is familiar from studies of other GN model variants, one would expect that the isoGN model also possesses bound states of several twisted kinks sitting at arbitrary
separations and whose mass is the sum of the individual kink masses. This can indeed be confirmed by using the duality between the isoGN and truncated isoNJL models.
For simplicity, we consider a bound state of two kinks. If one assumes that the two kinks have their isospin axes both in the (1,2)-plane, Eq.~(75) of Ref.~\cite{18}
shows that the isospin axis of the bound state remains in this plane everywhere. This enables us to construct the dual object. The general structure of the two-kink bound state in the isoGN model
will be
\begin{equation}.
\vec{S} = \frac{\vec{S}_0 + V_1 \vec{S}_1 + V_2 \vec{S}_2 + b_{12} V_1 V_2 \vec{S}_{12}}{1+V_1+V_2+b_{12}V_1V_2}
\label{E11}
\end{equation}
Here, the ${S}_i$ ($i=0,1,2,12$) are 3-vectors of length $M$ and
\begin{equation}
\vec{S}_0 \cdot \vec{S_i} = M^2 \cos (2\theta_i), \quad V_i = e^{2Mx\sin \theta_i}, \quad (i=1,2)
\label{E12}
\end{equation}
The physical interpretation of the vectors $\vec{S}_{i}$ is as follows: In isolation, soliton I interpolates between the vacua $\vec{S}_0$ and $\vec{S}_1$ with twist angle $\theta_1$, soliton II between $\vec{S}_0$ and $\vec{S}_2$ with
twist angle $\theta_2$.
Their bound state connects the vacua $\vec{S}_0$ and $\vec{S}_{12}$. If we choose the arbitrary spatial positions such that the solitons don't overlap, all four vectors $\vec{S}_i$ can be interpreted
as vacua, see Fig.~\ref{fig2} for the two possible orderings of the solitons. The most interesting question is: Given $\vec{S}_0,\vec{S}_1,\vec{S}_2$, what is $\vec{S}_{12}$, i.e., what is the composition law for twist?
We could answer this simply by transforming the explicit result of \cite{18} via duality, but a more instructive way is perhaps the following geometrical consideration.
Since the two plots in Fig.~\ref{fig2} are two different orderings of the same kinks, we must have
\begin{eqnarray}
\vec{S}_1\cdot \vec{S}_0 & = & \vec{S}_{12} \cdot \vec{S}_2,
\nonumber \\
\vec{S}_2\cdot \vec{S}_0 & = & \vec{S}_{12} \cdot \vec{S}_1.
\label{E16}
\end{eqnarray}
This merely expresses the fact that the twist angles are an intrinsic property of the kinks, independently of their relative positions in space.
It turns out that this is already sufficient to determine the unknown vector $\vec{S}_{12}$ up to a twofold discrete ambiguity. The solution which agrees 
with Eq.~(75) of Ref.~\cite{16} after the duality transformation is
\begin{eqnarray}
\vec{S}_{12} & = &  {\cal M}_{1,2}\vec{S}_0
\nonumber \\
{\cal M}_{1,2} & = & 1 - 2 \vec{e}_{1,2}\vec{e}_{1,2}^{\,T}
\nonumber \\
\vec{e}_{1,2} &  = & \frac{\vec{S}_1 - \vec{S}_2}{|\vec{S}_1-\vec{S}_2|}.
\label{E17}
\end{eqnarray}
Eq.~(\ref{E16}) is satisfied because all $\vec{S}_i$ have the same length, as one can easily check. The coefficient $b_{12}$ in (\ref{E11}) can also be
expressed in terms of the angles of the vectors $\vec{S}_{1,2}$ by the duality transformation, but we do not write down the complicated expression
which does not seem to have a simple geometrical interpretation. 

Finally, we note that time dependent solutions of the isoNJL model are also known explicitly. They include breathers and scattering problems of solitons or breathers.
If we try to transform the simplest example (scattering of two twisted kinks) into the isoGN model via duality, we find that even if $P_1=0$ initially, it does not stay 0 during the time evolution.
This seems to be unavoidable and prevents us from finding time dependent solutions of the isoGN model. The fact that static solutions can be written down in closed
form but time dependent ones apparently not is reminiscent of the massive GN model. In that case, it has been shown that integrability is lost when switching on the bare mass \cite{25}.
This may point to the fact that the isoGN model is not integrable, although we cannot rule out that time dependent solutions can be found by methods different from duality.

\section{Massive model}
\label{sect7}

Adding a bare mass term to the Lagrangian (\ref{I4}), we arrive at the 
massive isoGN model 
\begin{equation}
{\cal L}_{\rm isoGN} = \bar{\psi} \left(i \partial \!\!\!/- m_0 \right) \psi + \frac{g^2}{2} \left(\bar{\psi} \vec{\tau} \psi \right)^2.
\label{F1}
\end{equation}
The bare mass term breaks the discrete chiral symmetry, leaving SU(2) isospin intact.
In contrast to the other GN type models, the bare mass term yields a contribution to the mean field different from all terms
generated by the interaction and SSB. As we shall show, this has important consequences.

Let us consider the vacuum problem and the gap equation first. The HF Hamiltonian reads
\begin{equation}
H =- i  \gamma_5 \partial_x + \gamma^0 \left( m_0 + \vec{S} \cdot \vec{\tau} \right).
\label{F2}
\end{equation}
We diagonalize $H$ with constant $\vec{S}$ in momentum representation. The spectrum reveals two species of free, massive fermions
with masses split by $2m_0$,
\begin{equation}
M_{\pm} = |M \pm m_0|, \quad M=|\vec{S}|
\label{F4}
\end{equation}
and the vacuum energy density
\begin{eqnarray}
\frac{\cal E}{N_c} & = & - \int_{-\Lambda/2}^{\Lambda/2} \frac{dk}{2\pi} (\epsilon_+ + \epsilon_-) + \frac{M^2}{2 N_c g^2}
\nonumber \\
& = & - \frac{\Lambda^2}{4\pi} + \frac{1}{4\pi} \left( M_+^2 \ln \frac{M_+^2}{\Lambda^2} + M_-^2 \ln \frac{M_-^2}{\Lambda^2} - M_+^2- M_-^2 \right)  + \frac{M^2}{2 N_c g^2}.
\label{F5}
\end{eqnarray}
Minimizing with respect to $M$, we find the gap equation
\begin{equation}
\frac{2\pi}{N_c g^2} = 2 \ln \Lambda^2 - \ln (M_+^2 M_-^2) - \frac{m_0}{M} \ln \frac{M_+^2}{M_-^2}.
\label{F6}
\end{equation}
In the chiral limit ($m_0 \to 0$) this reduces to Eq.~(\ref{B4}). Alternatively, condition (\ref{F6}) could have been obtained from the self-consistency relation for the order parameter.
Upon using the gap equation to eliminate the coupling constant, the regularized vacuum energy density
becomes
\begin{equation}
\frac{\cal E}{N_c} =  - \frac{\Lambda^2}{4\pi} - \frac{M^2}{2\pi} + \frac{m_0^2}{2\pi} + \frac{m_0}{4\pi}\left( M_+ \ln M_+^2 - M_- \ln M_-^2 \right) - \frac{m_0^2}{\pi} \ln \Lambda.
\label{F7}
\end{equation}
Let us compare these findings with the corresponding results for the massive isoNJL model \cite{26}. There the gap equation was
\begin{equation}
\frac{2\pi}{N_c g^2} = 4 \left(\frac{m_0}{M}+1 \right) \ln \frac{\Lambda}{M} = 4 \left( \gamma + \ln \frac{\Lambda}{M} \right)
\label{F8}
\end{equation} 
with the ``confinement parameter" 
\begin{equation}
\gamma= \frac{\pi}{2N_c g^2} \frac{m_0}{M} = \frac{m_0}{M} \ln \frac{\Lambda}{M}.
\label{F9}
\end{equation}
In the case of the isoNJL model, one has to send $\Lambda \to \infty, m_0 \to 0$ keeping the physical parameter $\gamma$ constant. The bare mass $m_0$ cannot appear in any observable.
In the massive isoGN model, it does not seem to be necessary to renormalize the bare mass. The term $\sim m_0 \ln \Lambda$ in the gap equation (\ref{F8})
gets cancelled in the isoGN model when adding contributions from fermions with masses $M \pm m_0$. This suggests that the bare mass $m_0$ is a physical parameter
in the isoGN model. The new logarithmic divergence in the vacuum energy density (\ref{F7}) does not present any difficulty since
it is independent of the dynamical mass $M$, similar to the quadratic divergence. 

It is also instructive to look at the fate of the Goldstone bosons using RPA. In the NJL or isoNJL models with continuous chiral symmetries, the massless ``pions" acquire a mass
if one switches on the bare mass \cite{23}, obeying the Gell-Mann, Oakes, Renner relation \cite{27}. This is not expected here since the isospin symmetry is not broken explicitly by a bare mass term.
We have repeated the RPA calculation of Sec.~\ref{sect4}, using single particle energies and spinors appropriate to two species of fermions with masses $|M\pm m_0|$.
We find indeed again two massless scalar mesons. The marginally bound massive one disappears, similarly to what happens in the massive GN model.

Unfortunately, duality does not allow us to relate the massive versions of the isoGN and isoNJL models since the Dirac mass term $\sim \bar{\psi}\psi$ goes over into a term $\sim \bar{\psi} \tau_1 \psi $ .
Therefore we cannot say anything about the phase diagram or solitons of the massive isoGN model at this stage.

\section{Summary}
\label{sect8}
The first generation of (large $N_c$) four-fermion models in 1+1 dimensions comprises the GN and NJL models, featuring either a discrete or a continuous chiral symmetry.
Recently, there has been some interest in generalizing the NJL model by including isospin into the interaction. The resulting isoNJL model acquires a non-Abelian
chiral symmetry and is closer to the NJL model in 3+1 dimensions used as effective theory in strong interaction physics. The corresponding
generalization of the GN model, the isoGN model, has never been studied in 1+1 dimensions, to the best of our knowledge, although it 
has played a role in condensed matter physics in 2+1 dimensions (chiral Heisenberg Gross-Neveu model). The purpose of this work was to fill this gap.

To set the stage, we first determined the vacuum and the gap equation of the isoGN model. SSB of the discrete chiral symmetry and SU(2) isospin is found.
As verified using RPA, it is accompanied by the emergence of two massless bosons, matching the number of flat directions of the vacuum manifold (a 2-sphere).
Our most valuable tool however is a novel duality, mapping the isoGN model onto the isoNJL model. This was exploited to determine the phase diagram of the 
isoGN model as a function of temperature and two chemical potentials. 
Likewise, twisted kinks could be shown to exist in the isoGN model as well without any new effort. Static bound states of several kinks also carry over to the isoGN model.
The composition law for twist can be interpreted geometrically in isospin space.
Since the duality is between the isoGN model and an amputated version of the isoNJL model where the pseudoscalar  isovector mean field $\vec{P}$ is restricted to a plane, it has not been
possible to find time dependent solutions. This casts some doubts on the integrability of the isoGN model, unlike what is believed to hold for the GN, NJL and isoNJL models,
but we cannot rule out integrability at this stage.

Finally, we pointed out that adding a bare mass term to the isoGN model has a very different effect from all other models discussed.
The reason is the fact that there is no interaction in the scalar-isoscalar channel, so that the bare mass becomes a physical parameter without need for renormalization. 
The massless bosons remain massless if one switches on the bare fermion mass, in striking contrast to the usual scenario familiar from NJL type models.

In summary, we hope that the present study is of some pedagogical value, even if it does not have any phenomenological applications yet. It is based on
a natural generalization of the GN model and meant to fill an obvious gap in the otherwise well-explored family of GN type models of Table \ref{tab1}.

\end{document}